\documentstyle[12pt]{article}
\def\Mu{{\cal M}}
\def\cD{{\cal D}}
\begin{document}
\begin{center}
{\large\bf On fractal structure of quantum gravity and \\
relic radiation anisotropy}
\end{center}
\vskip48pt
\underline{M.V.~Altaisky}\footnote{E-mail: ALTAI@NUSUN.JINR.DUBNA.SU},
V.A.~Bednyakov and S.G.~Kovalenko \\
\vskip12pt
{\em Joint Institute for Nuclear Research, Dubna, 141980, \\
Russia}
\vskip2cm
\begin{abstract}
It is argued that large-scale ($>7^o$) cosmic microwave background
anisotropy detected in COBE cosmic experiment can be considered
as a trace of the quantum gravity fractal structure.
\end{abstract}

\topskip2in
\newpage

\section{Introduction}

It seems that a fractal structure is an intimate
property of the Universe. The superclusters (large clusters of galaxies
containing up to hundreds of thousands of galaxies) of size about
50 Mpc are separated by almost void space: the mean distance
between two superclusters is about 100 Mpc. Clusters of galaxies
(the typical cluster size is about 5 Mpc) containing hundreds of
galaxies are, in their turn, separated by voids about few Mpc. This
fractal hierarchy can be easily traced up to subnuclear scales
($10^{-13}$cm.).
\topskip5mm
Quantitatively, the large-scale fractal structure of the Universe
can be described in terms of the mass interior to the spherical
volume of certain radius $r$. The typical dependence, measured by
observing 21 cm hydrogen emmission of gas clouds moving around the
galaxy is
\begin{equation}
\Mu(r) \propto r^\alpha, \qquad \alpha\approx1, \label{R1}
\end{equation}
whereas a luminous mass associated with the light would supply
only $r^{-1/2}$. It is commonly accepted that an additional
mass in the form of non-luminous dark matter \cite{R1}. Since
$\alpha<3$ in the power law (\ref{R1}), we have a typical mass
distribution on a fractal set embedded in D=3 space.

On the other hand, one of the most important recent
developments in gravity theory was related to the
fractal based regularization
of quantum gravity \cite{KPZ}.
In view of this one may believe that a fractal structure is
a fundamental property of physical space-time itself.

In this note we interpret the COBE satellite data on the anisotropy
of cosmic microwave background radiation (CMBR) as a possible
manifestation of fractal structure of the Universe.

\section{On the discrete symmetry \newline in quantum gravity}

A regularization of two-dimensional quantum gravity, made by
V.~G.~Knizhnik, A.~M.~Polyakov and A.~B.~Zamolodchikov (KPZ),
comes from the fact that continuum formulation \cite{QG2D} and
the dynamical triangulation \cite{Boulatov} are equivalent.
On the basis of the Polyakov regularization procedure \cite{Polyakov},
where the position of the surface in the embedding space $X_\mu$
and the internal surface geometry ($g_{ab}$) are treated as
independent fields, one can construct a Nambu-like action
\begin{equation}
\begin{array}{ll}
S[X_\mu, g_{ab}] =& \\[8mm] = \frac{1}{2} \int_M
g_{ab} \frac{\partial X_\mu}{\partial\xi_a}
\frac{\partial X_\mu}{\partial\xi_b} \sqrt{\det g}\ d^2 \xi
+ \beta \int_M \sqrt{\det g}\ d^2 \xi + \label{nambu} \\[8mm]
+ \hbox{fermion terms} &
\end{array}
\end{equation}
where $\xi = (\xi_1,\xi_2)$ is the parametrization of the manifold $M$
defined by function $X_\mu = X_\mu(\xi)$.
This or a similar Nambu-Goto action usually stands in the string
functional integral taken with respect to both independent
fields $X$ and $g$.

In the dynamical triangulation \cite{Boulatov} of 2D quantum gravity,
 as well as in higher dimensional versions \cite{ambj},
the path integral over internal metric $g_{ab}$ is replaced by summation
of all the different types of surface configurations with given number
of triangles. For the sake of preserving reparametrization invariance
after discretization \cite{Boulatov}, the topology of the manifold $M$
is usually specified as the sphere $S^2$ \cite{Boulatov,KawRev}. The
partition function takes the form
\begin{equation}
Z(A) = \int_M \cD X \cD g \exp(-S), \label{PF}
\end{equation}
or its discrete counterpart \cite{KawRev}
\begin{equation}
Z_{reg}(A) = \sum_G Z_m(G) \delta_{Na^2,A},  \label{Zreg}
\end{equation}
where $A$ is the total area, N is the
number of equilateral triangles and $a^2$ is the area of an triangle.
The matter part of the partition function $Z_m(G)$ comes from the
fermion term of the KPZ lagrangian
\begin{equation}
{\cal L} = \bar\phi v^{\alpha a} \gamma^a \partial_\alpha \phi,
\end{equation}
where $v_{\alpha a}$ are ordinary "zweibeins".

Formally substituting functional integral (\ref{PF}) by
its discrete counterpart (\ref{Zreg}) we need to sum over
all possible triangulations of $S^2$. Practically, we are to
impose some additional conditions to avoid summation over
singular triangulations, i.~e.\ triangulations which include
links with coinciding ends. Referring the reader to
\cite{Boulatov,Kawamoto} for detailed study of triangulations
and fractal properties of related partition functions, we
shall concentrate on some of their properties
significant for the phenomenological applications.

{\em First.} The triangulation  procedure can be extended to an
$S^n$ sphere \cite{ambj}, which,  as a boundary of $(n+1)$-dimensional simplex,
can be divided into $n$-dimensional simplices.

{\em Second.} From the conformal invariance standpoint,
of all the subdivisions of $S^n$ the subdivision into
equilateral simplices is preferable.

{\em Third.} The whole partition function (\ref{PF}) is
related to the physical object, which is isotropic (in the
sense of having no preferable direction on $S^n$), but may
have a discrete symmetry group, and hence, have certain
distinguished correlation angles. For example, if we sum over
all possible triangulations of $S^2$ using equilateral
triangles, the correlations of any observables depending on
matter fields increase at angles
$0, \frac{2\pi}{3}, \frac{4\pi}{3}$ because of ${\rm Z}_3$
symmetry group. Similarly, the correlations should increase
at tetrahedron group angles when $S^3$ is considered.

{\em Fourth.} The two-dimensional quantum gravity can be
regarded as only the simplest case of extended object physics.
However, when reducing the physics from arbitrary
$n$-dimensional space to $(n-1)$ dimensions we
restrict $S^n$ triangulation with $n$-dimensional simplices
to $S^{n-1}$ triangulation with $(n-1)$ - dimensional ones,
because an $(n-1)$-dimensional simplex is a boundary of
an $n$-dimensional one. Thus, for the case of equilateral
symplices we should always have ${\rm Z}_3$-symmetry in
$D=2$, or tetrahedron symmetry in $D=3$.

\section{Discrete symmetry as a possible \newline source of relic
radiation anisotropy}

Let us consider the data \cite{CL1} on
relic radiation anisotropy. The
relic microwave radiation ($T = 2.73^o K$) was not significantly
affected by the late-stage processes in the Universe, that is why
its amplitudes depend mostly on the early Universe parameters.
It is worth to note, that the large scale anisotropy of relic
radiation, found in COBE and RELICT-1 experiments,
has a rather small value $\frac{\Delta T}{T}\sim 10^{-5}$, but a
high confidence level --- up to 90\%, including systematic errors
\cite{CL1,RELICT}.

Of course, the first aim of the observers in both COBE and RELICT
experiments was to measure the dipole and quadrupole components
of microwave background \cite{CL1} and to
test the existence of anomalous signal over the
mean background \cite{RELICT}. Basing on the
COBE experiment data, the autocorrelation function
\begin{equation}
C(\alpha) =
\langle \Delta T(\theta) \Delta T(\theta+\alpha) \rangle \label{cf}
\end{equation}
has been obtained.
Here $\alpha$ is the angle separation and $\theta$ is an angular
coordinate on certain two-dimensional plane. \marginpar{Fig. 1}


Qualitatively, the behavior of the relic signal autocorrelation
function (See fig.1)
is the following: it has a sharp maximum , it has another maximum
localized at $\alpha$ close to 120 degrees, and
it has two minimums at $60^o$ and $180^o$. (Maximum at $alpha$
close to $90^o$ is possibly related to quadrupole component of
CMBR and is less confidient \cite{CL7}).
The behavior of the
autocorrelation functions is just atmost the same for the data obtained at
frequencies 53 GHz and 90 GHz \cite{CL1}.

Correlation function (\ref{cf}) has been studied in
\cite{CL13} in connection with present cosmological models.
In particular, an attempt has been made to compare the COBE data
with certain Dark Matter (DM) models. This comparison does not
suit well. For instance, the relic density anisotropy given by Holtzman
model \cite{Holtzman} increases monotonously with $\alpha$ increasing
from 60 to 180 degrees \cite{CL13}.

Taking into account all the mentioned arguments, we interpret the regularities
of autocorrelation function (\ref{cf})
behavior, as a manifestation of ${\rm Z}_3$-symmetry.
The presense of ${\rm Z}_3$-symmetry does not imply $n$ preferable
directions in space here, instead we have a preferable separation
angle. It should be
mentioned that in COBE theoretical study \cite{CL13}
the best line fit
for autocorrelation function (\ref{cf}) was taken in the form
\begin{equation}
C(\alpha) =  A + B \cos \alpha + C^0_M \exp \left[
-\frac{\alpha^2}{2\sigma^2} \right], \label{cfa}
\end{equation}
though the locations of autocorrelation function
maximums at $0^o$ and $120^o$ and minimums at $60^o$ and $180^o$
suggest more direct parametrization
\begin{equation}
C(\alpha) =  A + B \cos 3\alpha + C^0_M \exp \left[
-\frac{\alpha^2}{2\sigma^2} \right] \label{cfa3}
\end{equation}

\section{On the flat-space limit of simplicial quantum gravity}

The simplest way to imagine how the distribution of relic radiation
with the simplicial symmetry could emerge from space-time geometry is
that of simplicial quantum gravity \cite{aj,am}. This theory enables one to
describe only pure gravity without matter
fields in a consistent way. Exact solution for matter coupling has been found
only for
two-dimensional case \cite{Kazakov,Boulatov}. For higher dimensions,
 if we want to describe the Nature as it is, we are to face
a lot of matter coupling problems. The question of our particular
concern should be the existence of flat-space continuous limit.
Hereinafter, we analyse the problems arising in $D>2$ simplicial
quantum gravity continuous limit and suggest a way to avoid them.

The very fact which, in our opinion, lead to KPZ regularization of
two-dimensional quantum gravity was the fractal nature of
dynamically triangulated
surface, rather than simplicial structure itself. (Investigation,
in some way similar to it, has been performed by Crane and Smolin
\cite{CrSm} without using triangulation at all.) That is why
we should expect some fractal structure which enables
 one to remove the divergences.

Indeed, in direct studies of quantum field theory models on fractal
space-time \cite{Ey}, as well as in studies devoted to fractal lattices
\cite{FrLat}, it has been shown that fractal sets, being scale-invariant,
are essentially relevant for treating by the renormalization group (RG)
technique.
Unfortunately, the price for
well defined scale properties is the lack of translation invariance.
This leads to the divergences. Till now this obstacle has not
been completely overcome.

As we consider the flat-space limit of (Euclidean) simplicial gravity,
we must pay attention to those fractal sets, which are suitable
for triangulation of an $S^n$ sphere. Thus, we consider {\em Sierpinski
hypergasket}, a generalization of the Sierpinski  gasket constructed
in two dimensions. Let us recall the construction procedure \cite{Ey}.
\marginpar{Fig. 2}

Partitioning the unit $d$-simplex in ${\rm R}^d$ into (d+2) sub-simplices
of edge-length $1/2$ one ($i$) removes the open central sub-simplex;
($ii$) repeates the operation with the $(d+1)$ closed sub-simplices.

Sierpinski gaskets obtained in this way can be used for triangulation of
an $S^n$-sphere. Their self-similarity is much
relevant to RG applications.  Their shortcomings are also evident. They are
not invariant under translations, even inside a single gasket, and they are
not dense in the embedding space.
That is why we are looking for the simplicial fractal set better in this
relation.

Let us modify the gasket generating procedure.
To clarify the consideration, let us imagine a unit simplex of
black color. On the first step of the recursive procedure we remove
the central open part of it, the central sub-simplex becomes "white",
and then --- here is the difference ---  repeat the procedure
with {\em all} $(d+2)$ sub-simplices. The generalization to "white"
pieces seems evident: the central part of each simplex reverses its color.
\marginpar{Fig. 3}

Since the numbers of "black" and "white" sub-simplices at the
$(k+1)$ stage of the recursive procedure are
\begin{equation}
\begin{array}{lcl}
n_{W}^{k+1} &=& (d+1) n_W^k + n_B^k \\
n_{B}^{k+1} &=& (d+1) n_B^k + n_W^k,
\end{array}
\end{equation}
for asymptotically large $k$ we obtain
$$ n_k \approx {1\over2} (d+2)^k$$
simplices of each color of $\delta = 2^{-k}$ edge-size.
The fractal dimension of the constructed set is
\begin{equation}
D = \frac{\log (d+2)}{\log 2},
\end{equation}
{\em rather than}
\begin{equation}
D = \frac{\log (d+1)}{\log 2},
\end{equation}
{for  Sierpinski gasket,}
But the scaling law is identical:
\begin{equation}
2^k \cdot {\cal G}(d) =  {\cal G}(d).
\end{equation}

As far as we know, there is no commonly accepted name to such a
set. Here, as the relation between black and white in the construction
is much like to ancient Chinese symbol of Yin and Yang, we can informally
call it Yin-Yang-gasket.

The geometrical properties of the above constructed set as a building
block for a piecewise approximation of an $S^n$ sphere ($n>2$) require
further investigation. Nevertheless, we can already mention the
 properties which could revive the simplicial quantum gravity
phenomenology.
The gasket is
{\em (i) simplicial, (ii) scale-invariant, (iii) homogeneous,
(iv) dense in embedding space}.

\section{Conclusion}

The data on relic radiation anisotropy obtained by
both RELICT and COBE groups are worth further deep investigation.
Nonetheless, even the results already obtained from data processing
seem to be in good agreement with the hypothesis of discrete
symmetry of space-time arising in fractal quantum gravity. Other
cosmological data, e.g. mass distribution, also do not contradict
either possible fractal structure of the Universe. It might be argued,
that both the tetrahedron symmetry, if found, and the fractal
structure of the visible Universe, can be regarded as an
argument for the existence of cosmic strings \cite{Kibble}.
Indeed, cosmic strings, as topological defects
which could be formed at a phase transition in the early Universe,
can have a number of cosmological applications. In particular, they
can form a network with a fractal structure having tetrahedron
symmetry \cite{string}.
Though the question, why cosmic strings must form the
tetrahedron structure still remains.

Therefore, some
new tests for possible discrete symmetry can be proposed. The
simplest among them are: ({\em i}) to test $\cos n\alpha, \quad n>1$
in (\ref{cfa}) for other ${\rm Z}_n$ groups, ({\em ii}) to use COBE and
RELICT data to
search for the tetrahedron or other essentially three-dimensional space
symmetry groups.
\vskip\baselineskip
\centerline{***}
\vskip\baselineskip

This work was supported in
part by the scientific program "Fundamental Metrology" of
Russian Ministry of Science.

The authors are also thankful do Dr. M.V.Sazhin for referring to COBE-group
results and useful discussions.

\newpage
\begin{center}
{\large \bf Figure captions}
\end{center}

\begin{itemize}
\item[{\bf Fig. 1}]
\vspace{12pt}
Correlation functions $C(\alpha)$, at various Gallactic lattitude cuts
for the 53MHz map. (Reprinted from \cite{CL1}).

\item[{\bf Fig. 2}]
\vspace{12pt}
Second stage of the ($2D$) Sierpinski gasket construction.

\item[{\bf Fig. 3}]
\vspace{12pt}
Second stage of the ($2D$) {\em Yin-Yang-gasket} construction.
The difference from fig.2 is shown in grey.
\end{itemize}
\end{document}